\documentclass[table, letterpaper, 10pt, conference]{ieeeconf} 
\IEEEoverridecommandlockouts 
\usepackage{graphicx}

\usepackage{indentfirst}
\usepackage{array,multirow}
\usepackage{fancyhdr}
\usepackage{float}
\usepackage{booktabs}
\usepackage{here}
\usepackage{comment}
\usepackage{siunitx}
\usepackage{algorithmicx,algorithm,algpseudocode}
\usepackage{amsmath,amssymb,amsfonts,bm,gensymb}
\usepackage{fontawesome5}
\usepackage{caption}
\usepackage{setspace}
\usepackage{xcolor}
\usepackage{cite}
 
\usepackage{svg}
\usepackage{media9}
\newcolumntype{R}{>{\centering\arraybackslash}m{0.8cm}}
\usepackage[normalem]{ulem}
\useunder{\uline}{\ul}{}
\usepackage{url,hyperref,lineno,microtype}

\def\BibTeX{{\rm B\kern-.05em{\sc i\kern-.025em b}\kern-.08em
  T\kern-.1667em\lower.7ex\hbox{E}\kern-.125emX}}

\usepackage{hhline} % used for the top-most horizontal line! in order to fix the top left corner cell empty
\usepackage{highlight}
\renewcommand{\opacity}{90}

\newcommand{\etal}{et~al.\ }

\title{\LARGE \bf
Where Do Passengers Gaze? Impact of Passengers' Personality Traits on Their Gaze Pattern Toward Pedestrians During APMV-Pedestrian Interactions with Diverse eHMIs
}

\author{
Hailong~Liu$^{1,*}$,~\IEEEmembership{Member,~IEEE,}  Zhe~Zeng$^{2}$ and Takahiro~Wada$^{1}$,~\IEEEmembership{Member,~IEEE}
\thanks{
$^{1}$Hailong~Liu and Takahiro~Wada are with the Graduate School of Science and Technology, Nara Institute of Science and Technology, 8916-5 Takayama-cho, Ikoma, Nara 630-0192, Japan.}
\thanks{
$^{2}$Zhe~Zeng is with Dept. Human Factors, Ulm University, Albert-Einstein-Allee 45, 89081, Ulm, Germany.}
\thanks{*CONTACT Hailong Liu. \faIcon[regular]{envelope}~:~{\tt\small liu.hailong@is.naist.jp}}
}

\begin{document}
\maketitle
\thispagestyle{empty}
\pagestyle{empty}

\begin{abstract}
Autonomous Personal Mobility Vehicles (APMVs) are designed to address the ``last-mile'' transportation challenge for everyone. 
When an APMV encounters a pedestrian, it uses an external Human-Machine Interface (eHMI) to negotiate road rights. 
Through this interaction, passengers are also passively exposed to the process. 
This study examines passengers' gaze behavior toward pedestrians during such interactions, focusing on whether passengers' personality traits influence their gaze patterns towards pedestrians when using different eHMI designs.
When using a visual-based eHMI, which caused passengers to struggle in perceiving the communication content, the results suggested that passengers with higher \textit{Neuroticism} scores, who were more sensitive to communication details, might seek cues from pedestrians' reactions.
In addition, a multimodal eHMI (visual and voice) using neutral voice did not significantly affect the gaze behavior of passengers toward pedestrians, regardless of personality traits.
In contrast, a multimodal eHMI using affective voice encouraged passengers with high \textit{Openness to Experience} scores to focus on pedestrians' heads.
In summary, this study revealed how different eHMI designs influence passengers' gaze behavior and highlighted the effects of personality traits on their gaze patterns toward pedestrians, providing new insights for personalized eHMI designs.
\end{abstract}

\section{INTRODUCTION}

% APMV

\subsection{Background}
Autonomous personal mobility vehicle (APMV) is a small autonomous vehicle designed to solve the last-mile problem for \textbf{everyone} (not only for the elderly or people with disabilities). 
APMV is equipped with SAE Level 3 to 5 automated driving systems and can drive autonomously in shared spaces (see Fig.~\ref{fig:intro}). 
In such driving scenarios, APMVs frequently encounter pedestrians and interact with them, even negotiating the right-of-way~\cite{liu2024_APMV_eHMI,liu2022implicit}.
To improve communication between APMVs and pedestrians, external human-machine interface~(eHMI), which is widely used in pedestrian and autonomous vehicle~(AV) research~\cite{lee2019autonomous,DeyGazePatterns2019,liu2025pre}, is also applied to APMVs~\cite{zhang2022understanding,liu2024_APMV_eHMI,zhang2024shared}.

\subsection{Challenges in eHMI Design for APMV Passengers}

\begin{figure}[t]

  \centering
  \includegraphics[width=1\linewidth]{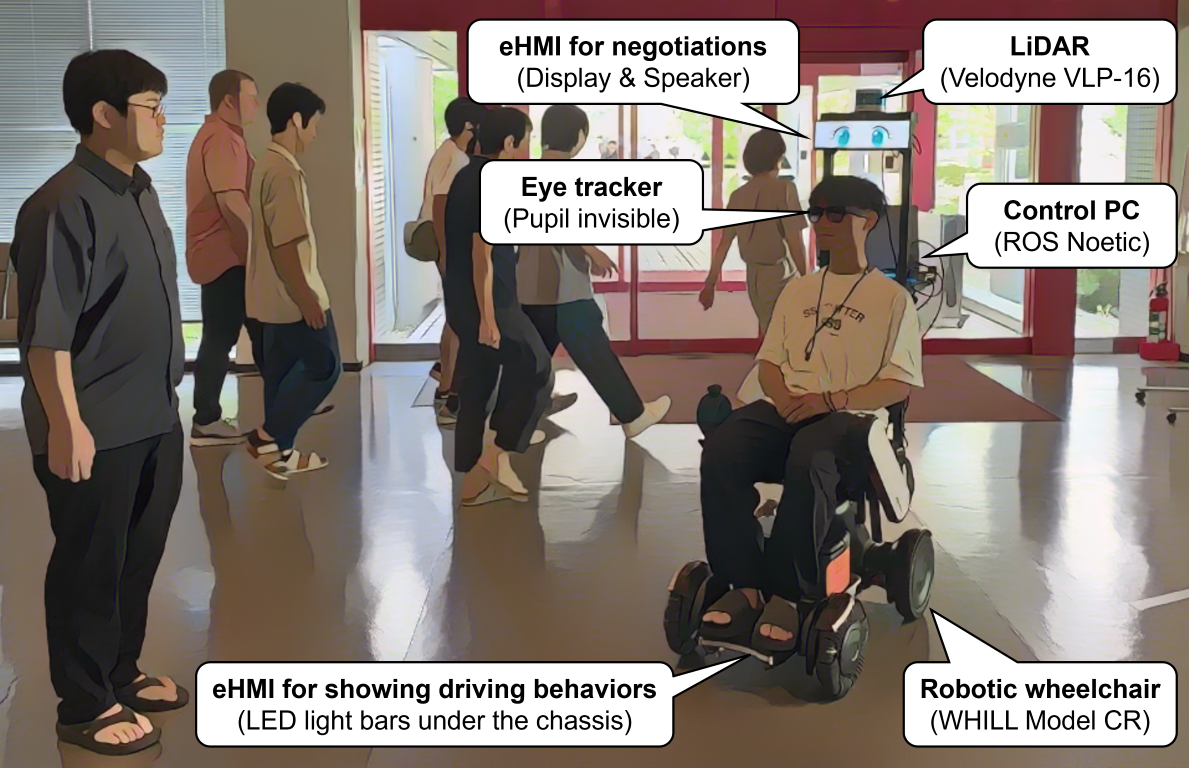}
  \caption{A passenger-carrying APMV equipped with an eHMI encounters a pedestrian in a shared space.}
  \label{fig:intro}
\end{figure}

Different from cars, APMVs often have an open-body design. 
Therefore, when an APMV uses eHMI to communicate with pedestrians, its passenger is also passively exposed to these interactions.
Based on Edward T. Hall's Proxemics Theory~\cite{hall1966hidden}, in pedestrian-APMV interactions, as shown in Fig.~\ref{fig:intro}, the distance between passengers and pedestrians is within the social distance range (1.2-3.6 meters). According to~\cite{hall1966hidden}, this range gives enough personal space and allows for clear communication between the passenger and the pedestrian APMV.
For this unique interaction among the pedestrian, the APMV, and the APMV passenger, it is important to consider both the efficiency of communication with pedestrians and the user experience of passengers when designing eHMI for APMVs~\cite{liu2024_APMV_eHMI}.

For this challenge, Zhang \etal \cite{zhang2024shared} suggested that when APMVs interact with pedestrians, the APMV should share its driving intention with both pedestrians and its passenger via an eHMI.
Thus, they proposed using a projector-type eHMI to display the driving path on the ground to show its driving intentions.
In our previous study, Liu \etal \cite{liu2024_APMV_eHMI} suggested a multimodal eHMI for APMV that uses LED lights to project colors under the APMV to show its driving intentions, and designed facial expressions and voices to communicate with pedestrians through a display and a speaker.
Moreover, Liu \etal \cite{liu2024_APMV_eHMI} analyzed subjective feelings of passengers during interactions between APMV and pedestrians. 
They found that passengers with different personality traits have different preferences for voice in eHMIs. Thus, they recommended considering passengers' personality differences to design eHMI of APMVs.
However, there is still a lack of clear guidelines on how to customize the eHMI design based on the personality of the passengers.

% In the study of human-to-human interactions, eye contact is also important for communication. 
% Specifically, people with different personality traits may exhibit different gaze behaviors during interactions.

% However, how the passengers perceive their environment and how personality traits influence their allocation of visual attention remain largely unknown. 

\subsection{Purpose}
In our previous study \cite{liu2024_APMV_eHMI}, we evaluated the subjective perspective of APMV passengers
interacted with pedestrians.
In this study, we continue on objective evaluation to explore how APMV passengers allocate their attention during encounters.
%we considered that APMV passengers seeking eye contact with pedestrians is key to expressing their willingness to engage in communication.
%Therefore, 
%This paper presents a different viewpoint, stating that in the interaction among the three entities, \ie APMV with eHMI, APMV passenger and pedestrian,
Therefore, this paper aims to explore whether the use of differently designed eHMIs on an APMV influences the gaze behaviors of passengers with different personality traits toward pedestrians' head during the APMV interactions with pedestrians.

% The research questions are:

% For example, when an eHMI uses an extroverted personality voice to interact with pedestrians, might it encourage extroverted passengers to engage in eye contact with pedestrians? 
% In contrast, could it cause introverted passengers to avoid eye contact with pedestrians due to feelings of awkwardness?

\section{RELATED WORKS}

Gaze is important for social interaction in daily life because people understand others' intentions through eye contact and use it to communicate~\cite{rauthmann2012eyes}.
In social interactions, gaze behavior, such as eye contact, also reflects different personality traits~\cite{libby1973personality,brooks1986effects}.

In the transportation field, gaze behavior has been widely studied in pedestrian crossing scenarios. 
For example, by analyzing pedestrians' gaze behavior to understand their crossing strategies, researchers have explored how pedestrians allocate visual attention during road-crossing tasks and how gaze behavior influences decision-making~\cite{zhao2023pedestrian}.
Furthermore, study~\cite{geruschat2003gaze} had examined the gaze behavior of normally sighted pedestrians when crossing various types of intersection safely.

Moreover, eye contact is one of gaze behavior, which serves as an important communication method between drivers, pedestrians, and cyclists~\cite{rasouli2017agreeing,li2021autonomous}.
Vulnerable road users, such as pedestrians and cyclists, often seek eye contact with the driver when encountering a vehicle~\cite{sahai2022crossing, de2021pedestrians}.
This eye contact from the driver indicates that the driver has recognized the pedestrian and assures the pedestrian's safety.
For example, the study \cite{onkhar2022effect} reported that eye contact between the driver and the pedestrian increased the percentage of participants indicating that it was safe to cross.

In research on human-personal mobility vehicle interactions, studies focusing on gaze analysis are limited.
The study~\cite{maekawa2019analysis} analyzed the gaze behavior of drivers of manually driven personal mobility vehicles. 
The researchers found that skilled drivers tend to focus on multiple potential risks in the driving environment at the same time.
Additionally, study~\cite{liu2022implicit} analyzed the gaze behavior of pedestrians when encountering an APMV. 
They found that when pedestrians do not understand the driving intentions of the APMV or feel threatened, the duration of their gaze toward the APMV increases significantly. 
They also suggested that the gaze behavior of pedestrians toward the APMV represents the pedestrians' attempt to required information about driving intentions from the APMV.

However, to the best of the authors' knowledge, there has been no research analyzing the gaze behavior of APMV passengers toward pedestrians during interaction between APMVs and pedestrians. Moreover, the impact of the APMV's eHMI designs on pedestrian gaze behavior in such interaction has yet to be explored.

% eye tracking 
%Understanding how humans use their eyes to gather information from their surroundings is fundamental to exploring the cognitive and perceptual processes that support ongoing behavior \cite{tatler2019eye}.

%The way people allocate visual attention in dynamic environments plays a critical role in guiding decision making, facilitating interactions, and ensuring safety.

% Wearable eye tracking glasses have emerged as a powerful tool to study these processes, offering insights into where and how people focus their attention in real-world human-vehicle interaction scenarios~\cite{liu2022implicit, geruschat2003gaze, DeyGazePatterns2019, de2021pedestrians}.

% In the field of human-vehicle interaction research, the analysis of gaze behavior has mostly focused on encounter scenarios between pedestrians and vehicles.
% Gaze plays a fundamental role in pedestrian-vehicle communication \cite{geruschat2003gaze, zhao2023pedestrian}. 
% Pedestrian safety and traffic efficiency largely depend on pedestrians' ability to accurately interpret the motion and intent of approaching vehicles.
% Traditionally, pedestrians make crossing decisions based on vehicle kinematics \cite{moore2019case}. 
% Moreover, pedestrians often rely on eye contact with drivers to communicate and express their intention~\cite{rasouli2017agreeing}. 

%gaze bahavior and personality.

\section{METHODOLOGY}

The experiment in this study is based on the experiment in the study~\cite{liu2024_APMV_eHMI}, which conducted a passenger-centered experiment using an APMV equipped with three types of eHMI designed for communication with pedestrians.
In the study~\cite{liu2024_APMV_eHMI}, data on the personality traits of passengers and subjective feelings were collected, and passengers' gaze behavior data was also collected during their interactions with pedestrians in the APMV.

Instead, in this paper, we focus on the gaze behavior of passengers associated with their personality traits regarding different eHMI conditions.
Therefore, in this study, we used the data on passengers' personality traits and gaze behavior from the experiment of the study~\cite{liu2024_APMV_eHMI} for analysis.
This experiment was approved by the Research Ethics Committee of Nara Institute of Science and Technology (NAIST) [No.~2022-I-55-1].

%This section briefly introduces the experimental design from \citep{liu2024_APMV_eHMI} and the causal discovery methods used in this paper.

\begin{figure*}[h!t]
  \centering
  \includegraphics[width=1\linewidth]{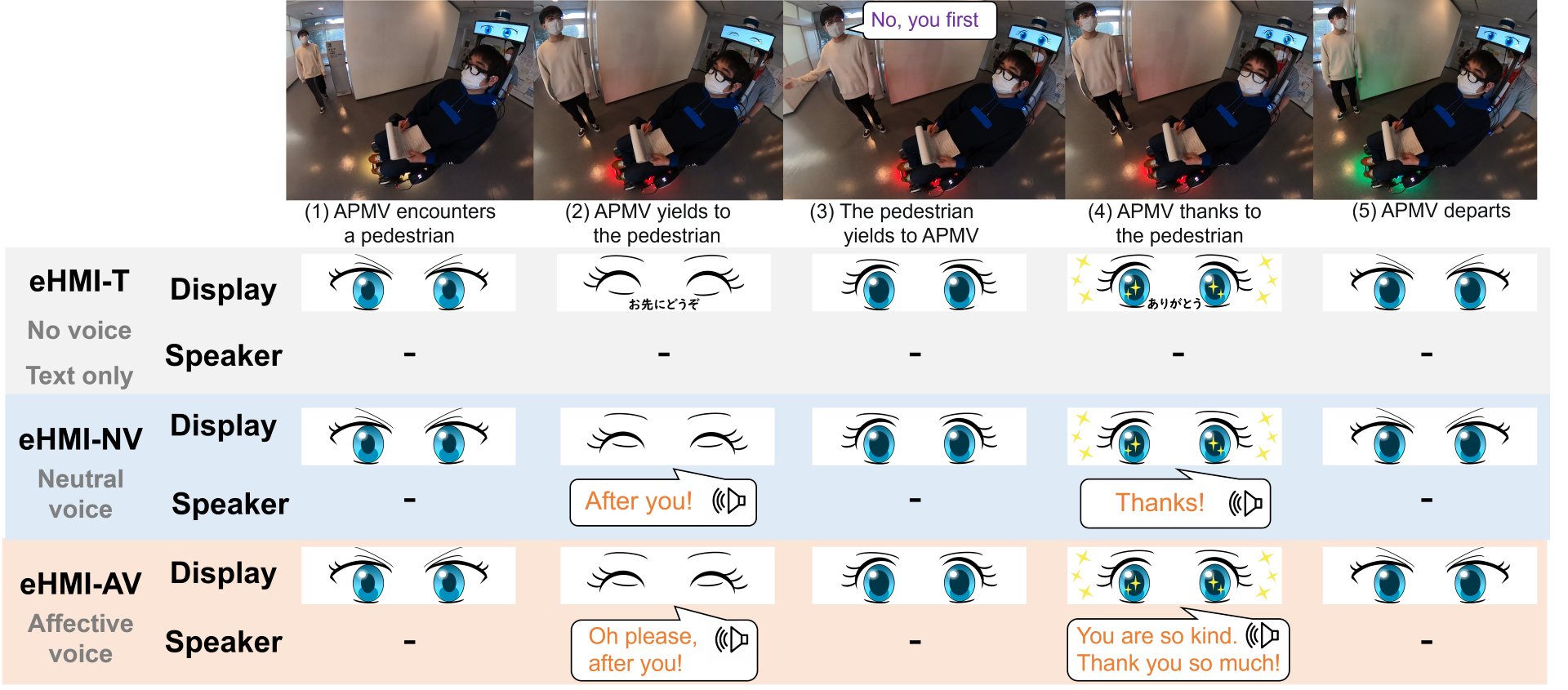}
  \caption{Three designed eHMIs for APMV negotiating with pedestrians~\cite{liu2024_APMV_eHMI}.}
  \label{fig:eHMIs}
\vspace{-2mm}
\end{figure*}

%\subsection{Equipment}

\subsection{Participants}
24 Japanese participants (12 males and 12 females, self-reported; 22 to 29 years with mean 23.9 years and standard deviation 1.7 years) participated in the ride experiment as APMV passengers.
All participants had no experience with APMV as well as eHMI.
Each participant took part in the experiment for one hour and received 1,000 JPY as a reward.

\subsection{Autonomous Personal Mobility Vehicle (APMV)}
As shown in Fig.~\ref{fig:intro}, an APMV is developed based on the {\it WHILL Model CR}, which is equipped with a LiDAR (Velodyne VLP-16) and a control computer.
An autonomous driving system (ADS) with SAE Level 3 was developed based on ROS Noetic~\footnote{ROS Noetic: \url{http://wiki.ros.org/noetic/}}.
This system uses LiDAR to detect obstacles, avoid them or stop, and automatically follow a predefined path.

%In this experiment, the maximum speed was set to 1 m/s (3.6 km/h) for safety reasons.
%Moreover, the APMV's stopping and starting during pedestrian interactions were controlled by an experimenter behind the scenes.

Even though the APMV has a Level 3 ADS and can autonomously drive on a pre-planned route, for safety reasons, the APMV's stopping and starting during pedestrian interactions were controlled by an experimenter behind the scenes.
Furthermore, the maximum speed was limited to 1 m/s (3.6 km/h) for the same safety reasons.

% \begin{figure*}[h!t]
%   \centering
%   %\includegraphics[width=1\linewidth]{Fig/APMV.png}
%   \includegraphics[width=1\linewidth]{Fig/APMV.png}
%   \caption{Autonomous Personal Mobility Vehicle (APMV) with external human-machine interface (eHMI).}
%   \label{fig:APMV}
%   %\vspace{-4mm}
% \end{figure*}

\subsection{Designs of External Human-Machine Interface (eHMI)}

To facilitate communication between the APMV and pedestrians, in this experiment, three eHMIs were designed: visual-based (eHMI-T), neural voice-based (eHMI-NV), and affective voice-based (eHMI-AV) as shown in Fig.~\ref{fig:eHMIs}. 
The eHMI-T only displays text with cartoon expressions on the display.
The eHMI-NV~\footnote{The voices of eHMI-NV and eHMI-AV can be found at \url{https://1drv.ms/f/s!AqdIEHyOvvX56E_YDn2VRzSeU1KK}\label{Nanami}} uses both cartoon expressions on the display and voice messages via the speaker to communicate with pedestrians.
It uses a neutral voice with brief content and an emotionless tone.
The eHMI-AV~\textsuperscript{\ref {Nanami}} has the same cartoon expression design as the eHMI-NV.
The difference is that eHMI-AV uses a more affective voice, with longer content and an emotional tone.

It should be noted that for eHMI-T, only pedestrians receive information, but passengers cannot.
In the cases of eHMI-NV and AV, the passenger on APMV can also get communication information through voice messages.

\subsection{Scenarios}

\begin{figure}[tb]
  \centering
  \includegraphics[width=1\linewidth]{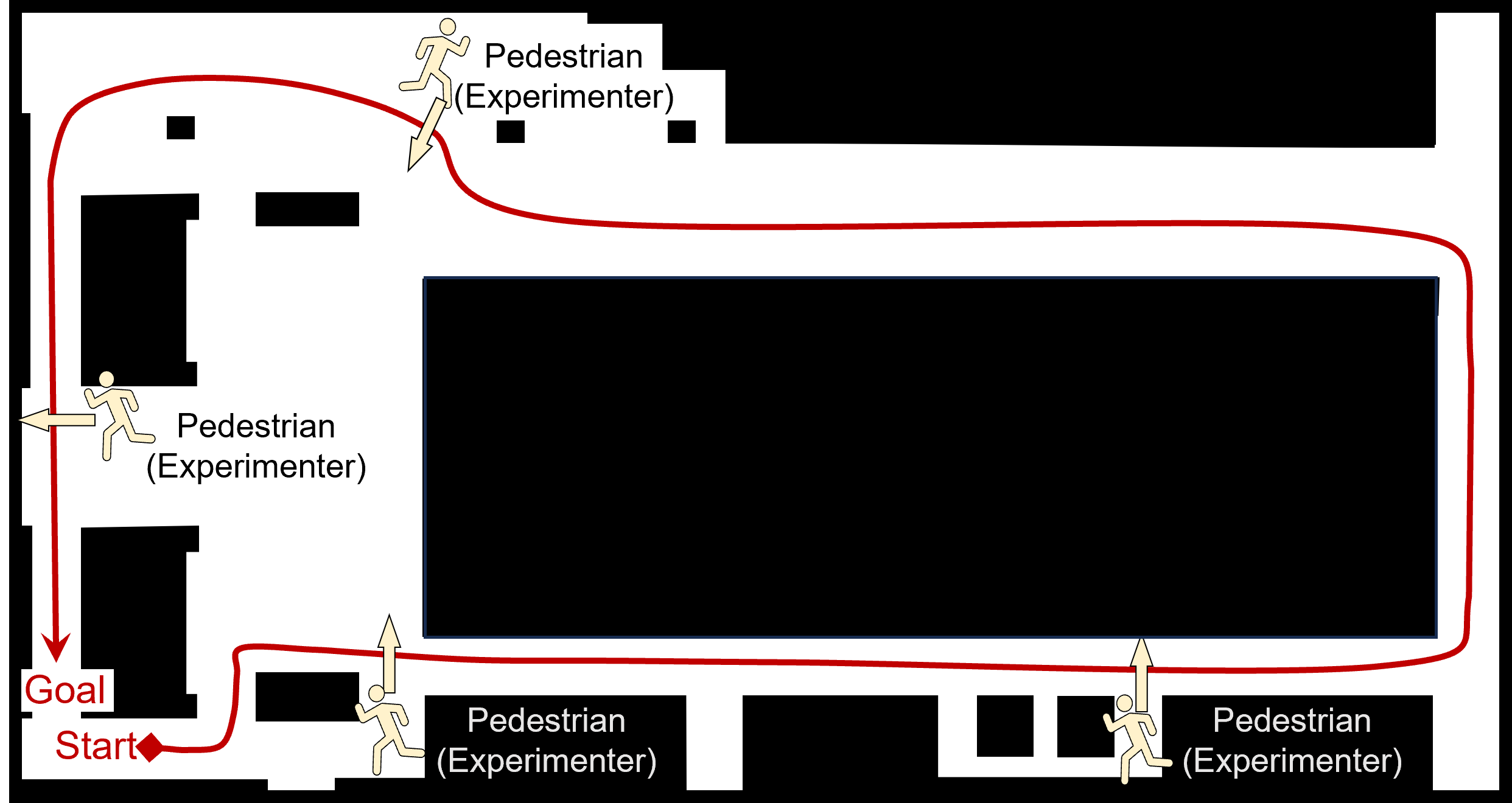}
  \caption{Experimental site and APMV driving route.}
  \label{fig:map}
  \vspace{-2mm}
\end{figure}

The experiment was carried out on the ground floor of a teaching building at NAIST, Japan (see Fig.~\ref{fig:map}).
Participants were seated as passengers on the APMV, which autonomously drove a circular corridor route (the red line in Fig.~\ref{fig:map}) three times, using a different eHMI in each round. 
During each round, the APMV encountered and negotiated with pedestrians played by experimenters four times.
To reduce order effects, the order of the three eHMIs is determined by a Latin square design (see Table~\ref{tab:balance}).

As shown in Fig.~\ref{fig:eHMIs}, in each encounter, APMV stops and uses the eHMI to convey a message indicating that it yields the right of way. 
After mutually yielding the right of way, the APMV uses the eHMI to express gratitude to the pedestrian and then prioritizes the passing.
During this interaction, as the pedestrian communicates with the eHMI, their gaze and facial orientation remain focused on the eHMI.

\begin{table}[t]
\centering
\footnotesize
\caption{The order of conditions balanced via Latin square design for the 24 participants (Rebuilt according to~\cite{liu2024_APMV_eHMI}.)}
\label{tab:balance}
\renewcommand{\arraystretch}{1}
\footnotesize
\setlength\tabcolsep{4.5pt}
\begin{tabular}{@{}crrr@{}}
\toprule
& \multicolumn{3}{c}{Experience orders of eHMIs} \\ \cmidrule(l){2-4} 
 Participants (N = 24) & \multicolumn{1}{c}{1st} & \multicolumn{1}{c}{2nd} & \multicolumn{1}{c}{3rd} \\ \midrule
Male = 2, Female = 2 & eHMI-T $\times$ 4 & eHMI-NV $\times$ 4 & eHMI-AV $\times$ 4 \\
Male = 2, Female = 2 & eHMI-T $\times$ 4 & eHMI-AV $\times$ 4 & eHMI-NV $\times$ 4 \\
Male = 2, Female = 2 & eHMI-NV $\times$ 4 & eHMI-T $\times$ 4& eHMI-AV $\times$ 4 \\
Male = 2, Female = 2& eHMI-NV $\times$ 4& eHMI-AV $\times$ 4& eHMI-T $\times$ 4\\
Male = 2, Female = 2 & eHMI-AV $\times$ 4 & eHMI-T $\times$ 4 & eHMI-NV $\times$ 4 \\
Male = 2, Female = 2 & eHMI-AV $\times$ 4 & eHMI-NV $\times$ 4& eHMI-T $\times$ 4\\ \bottomrule
\end{tabular}
%\vspace{-4mm}
\end{table}

\begin{figure}[t]
  \centering
  \includegraphics[width=0.75\linewidth]{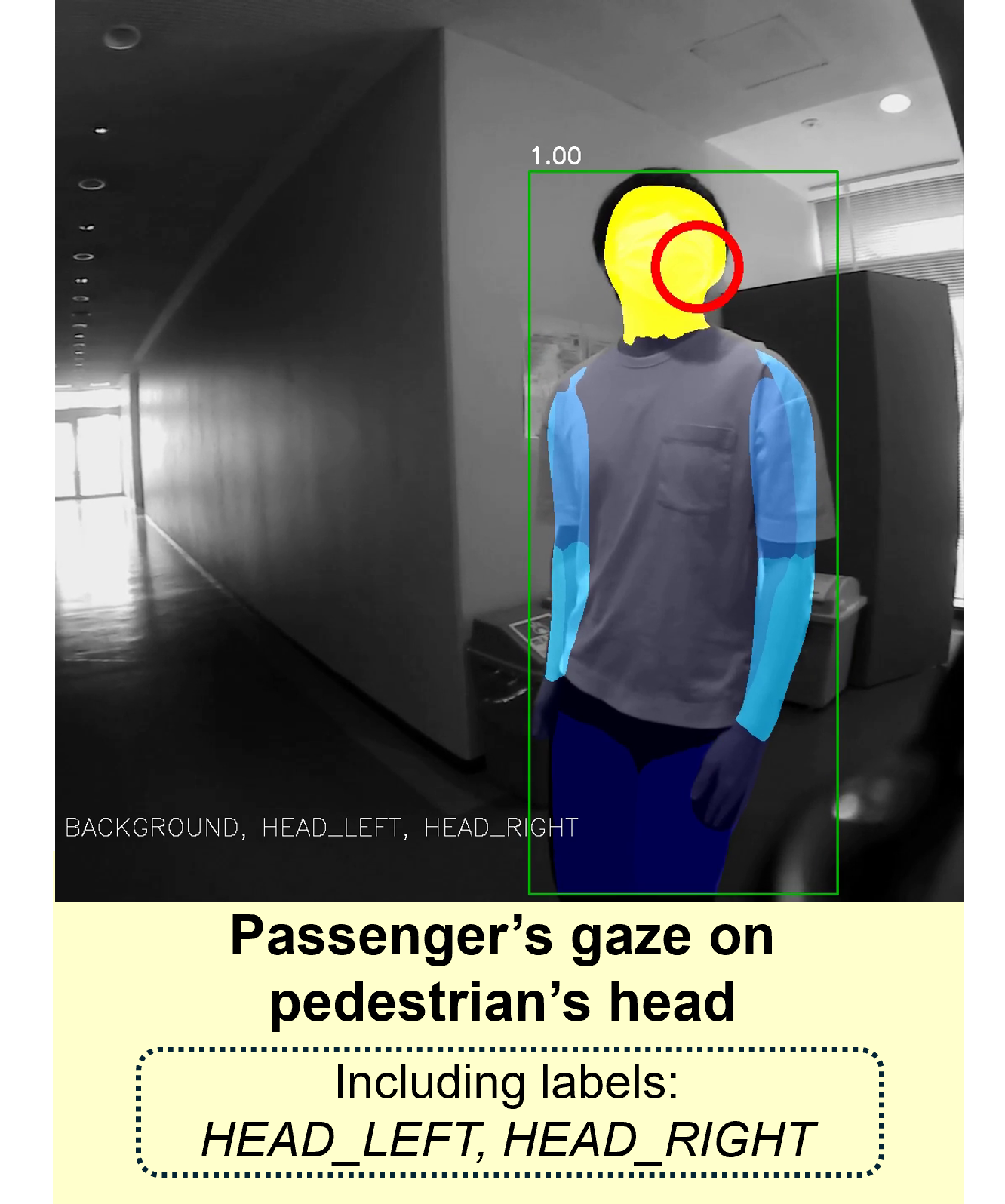}
  \caption{Area of interest (AOI) of pedestrian's head estimated by using the DensePose model from APMV passenger's gaze points}
  \label{fig:denspose_gaze}
\end{figure}

\subsection{Measurements}

\subsubsection{\textbf{Personality traits of passengers}}

The Big~5 personality model is commonly used to evaluate personality based on five traits: \textit{Openness to Experience}, \textit{Conscientiousness}, \textit{Extraversion}, \textit{Agreeableness}, and \textit{Neuroticism}~\cite{gosling2003very}.
To measure the Big~5 personality traits of passengers, the Japanese version of the Ten Item Personality Inventory (TIPI)~\cite{AtsushiOshio2012} was used.
According to \cite{AtsushiOshio2012}, the TIPI results can be calculated in the Big Five personality traits, with scores ranging from 1 to 14.

\subsubsection{\textbf{Passenger's gaze duration on pedestrian's head}}

The gaze behavior data of passengers during interactions between the APMV and pedestrians was recorded using Pupil Labs invisible eye-tracking glasses (sampling rate: 200 Hz).
The front-scene camera of the eye-tracking glasses recorded video with a resolution of $1088\times1080$ pixels in 30 Hz, providing a first-person perspective of the participants.

%how to use densepose
We use an image-based human detection model called DensePose~\cite{Guler2018DensePose}.
DensePose can recognize and label different parts of the human body from the recored images via the front-scene camera and assign a corresponding body part label to each pixel.
By combining passenger gaze data, we can calculate whether the passenger looked at the pedestrian during the interaction between the APMV and the pedestrian, and which part of the pedestrian's body was looked at.

As shown in Fig.~\ref{fig:denspose_gaze},
an Area of Interest (AOI) is defined to analyze passengers' gaze behavior toward the pedestrians' head.
More specifically, the DensePose labels ``\textit{HEAD\_LEFT}'' and ``\textit{HEAD\_RIGHT}'' used to define the AOI are shown in Fig.~\ref{fig:denspose_gaze}.
The red circle, with a diameter of 50 pixels, represents the gaze position. 
The areas covered by the red circle that overlap with the AOI regions are classified as gazes on it. 
As shown in Fig.~\ref{fig:denspose_gaze}, if the DensePose prediction output contains multiple prediction results, any result that includes either ``\textit{HEAD\_LEFT}'' and ``\textit{HEAD\_RIGHT}'' is also defined as the gaze on the head.

%In particular, if the red circle simultaneously includes HD, BD, and BK, it is classified as a gaze on HD. Similarly, if the red circle includes both BD and BK, it is classified as a gaze on BD.

%The duration of each interaction between the APMV and the pedestrians is defined within 15 seconds.
%During this 15-second period, the gaze duration for each AOI is defined as the total time spent gaze on the corresponding AOI.

% In which, a fixation is defined as the duration of gaze stability set between 80 and 220 milliseconds with max dispersion of 1.5 angle. 

% \section{Experimental design}
% potential dependent variables: gaze duration, average fixation duration, first fixation to AOI

% AOIs: the pedestrian as one AOI, or based on body part (head, and body).

% Participants who feel less awkward will tend to look at the head AOI longer and more frequently compared to those who feel more awkward (paper might relate to it - Eye contact avoidance in crowds: A large wearable eye-tracking study). 

% \begin{figure*}[t]
%   \centering
%   \includegraphics[width=0.75\linewidth]{Fig/denspose_gaze.pdf}
%   \caption{Areas of interest (AOI) of pedestrian's head estimated by using the DensePose model from APMV passenger's gaze points}
%   \label{fig:denspose_gaze}
% \end{figure*}

\section{RESULTS}

\subsection{Big~5 personality traits}

Distributions of the participants' Big~5 personality traits are shown in Table~\ref{tab:big5}.
The mean and standard deviation (std.) for each of Big~5 personality traits collected from the 24 Japanese participants in this experiment, as well as in the reference study~\cite{AtsushiOshio2012} for 902 Japanese participants are shown in Table~\ref{tab:big5}.
Using the distribution from~\cite{AtsushiOshio2012} as a reference, the \textit{Extraversion}, \textit{Agreeableness}, \textit{Conscientiousness}, and \textit{Openness to Experience} scores in this experiment are slightly higher, while \textit{Neuroticism} is slightly lower.

% \begin{figure*}[ht]
% \centering
% \includegraphics[width=1\linewidth]{Fig/Big5.pdf}
% %\vspace{-8mm}
% \caption{Distributions of Big~5 personality traits (ranging from 1 to 14) observed from 24 participants.}
% \label{fig:Big5}
% \end{figure*}

\begin{table}[ht]
\centering
\footnotesize
\setlength\tabcolsep{2pt}
\caption{The distribution of each Big Five personality trait measured in this study and its comparison with~\cite{AtsushiOshio2012}.}
\label{tab:big5}
\begin{tabular}{@{}lcccc@{}}
\toprule
& \multicolumn{2}{c}{\begin{tabular}[c]{@{}c@{}}This study\\ Japanese \textit{N} = 24\\ (male: 12, female: 12)\end{tabular}} & \multicolumn{2}{c}{\begin{tabular}[c]{@{}c@{}}Oshio 2021 \cite{AtsushiOshio2012}\\Japanese \textit{N} = 902 \\ (male: 376, female: 526)\end{tabular}} \\ \cmidrule(r){2-3} \cmidrule(l){4-5}
Big five traits & Mean & Std. & Mean & Std. \\ \cmidrule(r){1-1}\cmidrule(r){2-3} \cmidrule(l){4-5}
Extraversion & 9.54 & 2.89 & 7.83 & 2.97 \\
Agreeableness & 10.08 & 2.26 & 9.48 & 2.16 \\
Conscientiousness & 7.04 & 2.68 & 6.14 & 2.41 \\
Neuroticism & 7.50 & 2.69 & 9.21 & 2.48 \\
Openness to Experience & 9.75 & 2.69 & 8.03 & 2.48 \\ \bottomrule 
\end{tabular}
%\vspace{-2mm}
\end{table}

\subsection{Gaze duration under eHMI conditions}

\begin{figure}[t]
  \centering
  \includegraphics[width=0.74\linewidth]{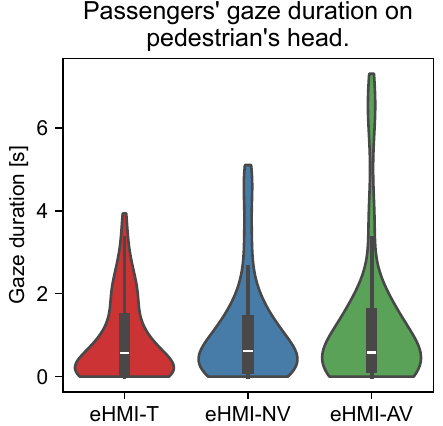}
  \caption{Passengers' gaze duration on pedestrians' head under eHMI conditions}
  \label{fig:Gaze_duration}
  \vspace{-4mm}
\end{figure}

When APMVs negotiate with pedestrians via three types of eHMIs, passengers' gaze durations on pedestrians' head are shown by violin plots in Fig.~\ref{fig:Gaze_duration}.
A Shapiro-Wilk test reported that the distribution of passengers' gaze durations on pedestrians' head departed significantly from normality under conditions of eHMI-T ($W=0.831, p<0.001$), eHMI-NV ($W=0.774, p<0.001$) and eHMI-AV ($W=0.675, p<0.001$).
Therefore, a Friedman F test was performed and showed that there was no significant differences in the gaze durations on pedestrians' head under the three eHMI conditions ($F=3.028, p=0.060$).

% \begin{figure*}[h!t]
%   \centering
%   \includegraphics[width=0.9\linewidth]{Fig/Gaze_eHMI.pdf}
%   \caption{Gaze duration under eHMI conditions}
%   \label{fig:Gaze_duration}
% \end{figure*}

\subsection{Pairwise Pearson Correlations Analysis}

Pairwise pearson correlation analysis to explore the linear relationship between each personality traits and gaze behavior under different eHMI conditions.

Tables~\ref{tab:P_cor_eHMI-T}, \ref{tab:P_cor_eHMI-NV} and \ref{tab:P_cor_eHMI-AV} show the results of the pairwise correlation analysis between passengers' personality traits and gaze durations on pedestrians' head during APMV interactions with pedestrians using the eHMI-T, eHMI-NV, and eHMI-AV, respectively.

The above results show that when the APMV uses the eHMI-T, there is a significant weak positive correlation between participants' \textit{Neuroticism} scores and their gaze duration on head ($r = 0.206, p = 0.047$).

When using eHMI-NV, no significant correlation was found between the Big Five personality traits of the passengers and their gaze duration on pedestrians' head.

When the APMV uses the eHMI-AV, a significant weak positive correlation was found between passengers' \textit{Openness to Experience} scores and their gaze duration on pedestrians' head ($r = 0.220, p = 0.032$). 
%Additionally, passengers' \textit{Agreeableness} scores showed a significant weak negative correlation with their gaze duration on BD ($r = -0.251, p = 0.014$), while their \textit{Neuroticism} scores showed a significant weak positive correlation with their gaze duration on BD ($r = 0.214, p = 0.037$).

\begin{table*}[h]
  \centering
    \caption{Pearson correlations between passengers' personality traits and their gaze duration on pedestrians' head during APMV interactions with pedestrians using the eHMI-T.}
  \label{tab:P_cor_eHMI-T}
  \footnotesize
  \renewcommand{\arraystretch}{0.9}
  \setlength\tabcolsep{13pt}
\begin{tabular}{llrrllrr}
\toprule
\multicolumn{1}{c}{ X }          & \multicolumn{1}{c}{Y }       &  \textit{n}$^{\dagger}$ &   \multicolumn{1}{c}{\textit{r}} & \multicolumn{1}{c}{CI~95\%}    & \multicolumn{1}{c}{ \textit{p}} & \multicolumn{1}{c}{ BF10} &  \multicolumn{1}{c}{\textit{power}} \\
\midrule
Extraversion      & Gaze duration on head   & 93 & -0.168 & [-0.36 ~0.04] &  0.107 & 0.467 &  0.367 \\
 Agreeableness     & Gaze duration on head   & 93 & -0.069 & [-0.27 ~0.14] &  0.512 & 0.160 &  0.101 \\
 Conscientiousness   & Gaze duration on head   & 93 & -0.003 & [-0.21 ~0.20] &  0.978 & 0.130 &  0.050 \\
 Neuroticism      & Gaze duration on head   & 93 & 0.206 & [~0.00 ~0.39]  &  0.047 * & 0.901 &  0.514 \\
 Openness to Experience & Gaze duration on head   & 93 & -0.048 & [-0.25 ~0.16] &  0.647 & 0.144 &  0.074 \\ 
 % \midrule
 % Extraversion      & Gaze duration on BD    & 93 & -0.093 & [-0.29 ~0.11] &  0.373 & 0.191 &  0.145 \\
 % Agreeableness     & Gaze duration on BD    & 93 & -0.029 & [-0.23 ~0.18] &  0.784 & 0.134 &  0.058 \\
 % Conscientiousness   & Gaze duration on BD    & 93 & -0.086 & [-0.29 ~0.12] &  0.410 & 0.181 &  0.131 \\
 % Neuroticism      & Gaze duration on BD    & 93 & 0.042 & [-0.16 ~0.24] &  0.686 & 0.140 &  0.069 \\
 % Openness to Experience & Gaze duration on BD    & 93 & -0.046 & [-0.25 ~0.16] &  0.661 & 0.143 &  0.072 \\ \midrule
 % Extraversion      & Gaze duration on BK & 93 & 0.187 & [-0.02 ~0.38] &  0.073 & 0.632 &  0.438 \\
 % Agreeableness     & Gaze duration on BK & 93 & 0.072 & [-0.13 ~0.27] &  0.495 & 0.163 &  0.105 \\
 % Conscientiousness   & Gaze duration on BK & 93 & 0.047 & [-0.16 ~0.25] &  0.657 & 0.143 &  0.073 \\
 % Neuroticism      & Gaze duration on BK & 93 & -0.192 & [-0.38 ~0.01] &  0.065 & 0.695 &  0.459 \\
 % Openness to Experience & Gaze duration on BK & 93 & 0.063 & [-0.14 ~0.26] &  0.546 & 0.155 &  0.093 \\
\bottomrule
\multicolumn{8}{l}{ \footnotesize \renewcommand{\arraystretch}{1} \begin{tabular}[c]{@{}l@{}} \textit{n}: number of trials. *:$p<0.05$. %; HD: Head; BD: Body w/o head; BK: Background.\\ 
~ ~ $^{\dagger}$: Gaze data from 3 out of 4 trials for one participant were defective, so these data of 3 trials were excluded.\\
%*:$p<0.05$.
 \end{tabular}}
 \end{tabular}
\vspace{-2mm}
\end{table*}

\begin{table*}[h]
  \centering
    \caption{Pearson correlations between passengers' personality traits and their gaze duration on pedestrians' head during APMV interactions with pedestrians using the eHMI-NV.}
  \label{tab:P_cor_eHMI-NV}
  \footnotesize
    \renewcommand{\arraystretch}{0.9}
    \setlength\tabcolsep{13pt}
\begin{tabular}{llrrllrr}
\toprule
\multicolumn{1}{c}{ X }          & \multicolumn{1}{c}{Y }       &  \textit{n} &   \multicolumn{1}{c}{\textit{r}} & \multicolumn{1}{c}{CI~95\%}    & \multicolumn{1}{c}{ \textit{p}} & \multicolumn{1}{c}{ BF10} &  \multicolumn{1}{c}{\textit{power}} \\
\midrule
 Extraversion      & Gaze duration on head   & 96 & 0.096 & [-0.11 ~0.29] &  0.353 & 0.195 &  0.154 \\
 Agreeableness     & Gaze duration on head   & 96 & -0.095 & [-0.29 ~0.11] &  0.359 & 0.193 &  0.151 \\
 Conscientiousness   & Gaze duration on head   & 96 & -0.028 & [-0.23 ~0.17] &  0.785 & 0.132 &  0.058 \\
Neuroticism      & Gaze duration on head   & 96 & -0.001 & [-0.20 ~0.20]  &  0.991 & 0.128 &  0.050 \\
 Openness to Experience & Gaze duration on head   & 96 & 0.193 & [-0.01 ~0.38] &  0.059 & 0.737 &  0.475 \\ 
 % \midrule
 % Extraversion      & Gaze duration on BD    & 96 & -0.055 & [-0.25 ~0.15] &  0.596 & 0.147 &  0.083 \\
 % Agreeableness     & Gaze duration on BD    & 96 & -0.053 & [-0.25 ~0.15] &  0.607 & 0.145 &  0.081 \\
 % Conscientiousness   & Gaze duration on BD    & 96 & -0.055 & [-0.25 ~0.15] &  0.594 & 0.147 &  0.083 \\
 % Neuroticism      & Gaze duration on BD    & 96 & 0.008 & [-0.19 ~0.21] &  0.939 & 0.128 &  0.051 \\
 % Openness to Experience & Gaze duration on BD    & 96 & 0.020 & [-0.18 ~0.22] &  0.850 & 0.130 &  0.054 \\ \midrule
 % Extraversion      & Gaze duration on BK & 96 & -0.031 & [-0.23 ~0.17] &  0.762 & 0.133 &  0.060 \\
 % Agreeableness     & Gaze duration on BK & 96 & 0.104 & [-0.10 ~0.30]  &  0.312 & 0.211 &  0.173 \\
 % Conscientiousness   & Gaze duration on BK & 96 & 0.058 & [-0.14 ~0.26] &  0.575 & 0.149 &  0.087 \\
 % Neuroticism      & Gaze duration on BK & 96 & -0.005 & [-0.20 ~0.20]  &  0.965 & 0.128 &  0.050 \\
 % Openness to Experience & Gaze duration on BK & 96 & -0.152 & [-0.34 ~0.05] &  0.139 & 0.375 &  0.317 \\
\bottomrule
\multicolumn{8}{l}{ \footnotesize \renewcommand{\arraystretch}{1}
 \begin{tabular}[c]{@{}l@{}} \textit{n}: number of trials. %; HD: Head; BD: Body w/o head; BK: Background.\\ 
 \end{tabular}}
\end{tabular}
\vspace{-2mm}
\end{table*}

\begin{table*}[h]
  \centering
    \caption{Pearson correlations between passengers' personality traits and their gaze duration on pedestrians' head during APMV interactions with pedestrians using the eHMI-AV.}
  \label{tab:P_cor_eHMI-AV}
  \footnotesize
    \renewcommand{\arraystretch}{0.9}
    \setlength\tabcolsep{13pt}
\begin{tabular}{llrrllrr}
\toprule
\multicolumn{1}{c}{ X }          & \multicolumn{1}{c}{Y }       &  \textit{n} &   \multicolumn{1}{c}{\textit{r}} & \multicolumn{1}{c}{CI~95\%}    & \multicolumn{1}{c}{ \textit{p}} & \multicolumn{1}{c}{ BF10} &  \multicolumn{1}{c}{\textit{power}} \\
\midrule
 Extraversion      & Gaze duration on head   & 96 & 0.109 & [-0.09 ~0.30] &  0.291 & 0.221 &  0.185 \\
 Agreeableness     & Gaze duration on head   & 96 & -0.020 & [-0.22 ~0.18] &  0.845 & 0.130 &  0.054 \\
 Conscientiousness   & Gaze duration on head   & 96 & 0.134 & [-0.07 ~0.33] &  0.192 & 0.295 &  0.258 \\
 Neuroticism      & Gaze duration on head   & 96 & 0.058 & [-0.14 ~0.26] &  0.574 & 0.149 &  0.087 \\
 Openness to Experience & Gaze duration on head   & 96 & 0.220 & [~0.02 ~0.40]  &  0.032 * & 1.244 &  0.581 \\ 
 % \midrule
 %  Extraversion      & Gaze duration on BD    & 96 & -0.073 & [-0.27 ~0.13] &  0.479 & 0.163 &  0.109 \\
 % Agreeableness     & Gaze duration on BD    & 96 & -0.251 & [-0.43 -0.05] &  0.014 * & 2.547 &  0.700 \\
 % Conscientiousness   & Gaze duration on BD    & 96 & -0.141 & [-0.33 ~0.06] &  0.171 & 0.320 &  0.279 \\
 %  Neuroticism      & Gaze duration on BD    & 96 & 0.214 & [~0.01 ~0.40]  &  0.037 * & 1.096 &  0.556 \\
 %  Openness to Experience & Gaze duration on BD    & 96 & -0.198 & [-0.38 ~0.00] &  0.053 & 0.800 &  0.492 \\ \midrule
 %   Extraversion      & Gaze duration on BK & 96 & -0.064 & [-0.26 ~0.14] &  0.534 & 0.154 &  0.095 \\
 % Agreeableness     & Gaze duration on BK & 96 & 0.143 & [-0.06 ~0.33] &  0.165 & 0.330 &  0.286 \\
 % Conscientiousness   & Gaze duration on BK & 96 & -0.054 & [-0.25 ~0.15] &  0.600 & 0.146 &  0.082 \\
 % Neuroticism      & Gaze duration on BK & 96 & -0.159 & [-0.35 ~0.04] &  0.121 & 0.417 &  0.344 \\
 % Openness to Experience & Gaze duration on BK & 96 & -0.105 & [-0.30 ~0.10]  &  0.310 & 0.212 &  0.174 \\
\bottomrule
\multicolumn{8}{l}{ \footnotesize \renewcommand{\arraystretch}{1}
 \begin{tabular}[c]{@{}l@{}} \textit{n}: number of trials. %; HD: Head; BD: Body w/o head; BK: Background.\\ 
*:$p<0.05$.
 \end{tabular}}
\end{tabular}
\vspace{-5mm}
\end{table*}

\section{DISCUSSION}

\subsection{Effects of eHMI-T on passenger gaze behaviors}
From the pairwise correlation analysis results shown in Table~\ref{tab:P_cor_eHMI-T}, when the APMV communicates with pedestrians using eHMI-T, there was only a significant correlation relationship was observed between the passengers' \textit{Neuroticism} scores and their gaze duration on the pedestrians' head with a positive correlation.
This result suggests that passengers with lower \textit{Neuroticism} tended to avert their gaze from the pedestrian's head while immersed in the interaction with the pedestrian and the silent eHMI-T, whereas passengers with higher \textit{Neuroticism} tended to allocate longer visual attention to the pedestrian's head.

However, studies~\cite{uusberg2015eye,jensen2016personality} suggested that people with a high \textit{Neuroticism} score tended to look away when co-communicators looked them in the eyes during a conversation or other social situations.
We considered that this difference arises because the studies~\cite{uusberg2015eye,jensen2016personality} focused on face-to-face interactions, where mutual eye contact was more likely to occur.
In our experiment, the pedestrian primarily interacted with the eHMI and did not look at the passenger throughout the process.
Moreover, since passengers could not directly access the communication content from the eHMI-T when the APMV interacted with pedestrians (study~\cite{liu2024_APMV_eHMI} also highlighted this issue via subjective evaluations), they had to infer it based on the pedestrians' reactions.
Given this uncertainty, we considered that passengers with higher neuroticism scores might remain more vigilant in such an ambiguous communication environment.
Studies~\cite{bradley2000covert, perlman2009individual} provided some thread showing that people with anxious personalities, such as high \textit{Neuroticism}, exhibit increased vigilance toward potentially threatening stimuli.

Therefore, we considered that this heightened vigilance could lead them to spend more time gazing on the pedestrian's head in an effort to carefully interpret the interaction between the eHMI-T and the pedestrian.

\subsection{Effects of eHMI-NV on passenger gaze behaviors}

From the results of the pairwise correlation analysis shown in Table~\ref{tab:P_cor_eHMI-NV}, when using eHMI-NV, no significant correlation was found between passengers' Big Five personality traits and their gaze behavior. 
This result suggested that the neutral communication style of eHMI-NV provided a relatively consistent experience for passengers with different personality traits, and it was difficult to affect the gaze behavior of passengers with specific personality traits.

\subsection{Effects of eHMI-AV on passenger gaze behaviors}

When using eHMI-AV, we found the \textit{Openness to Experience} scores of the passengers positively correlated with their gaze duration on head as shown in Table~\ref{tab:P_cor_eHMI-AV}.
Generally, people with high \textit{Openness to Experience} are more inclined to embrace new experiences, exhibiting heightened curiosity and a desire for exploration. 
Therefore, the rich emotional voice provided by eHMI-AV might offer a novel experience for passengers, especially those with high \textit{Openness to Experience} passengers, as they might be more willing to engage in the interaction between the APMV and pedestrians.
This aligns with the findings of study~\cite{zohoorian2022willingness} which reported a correlation between willingness to communicate and \textit{Openness to Experience}.
In addition, this also can be attributed to the fact that participants with high \textit{Openness to Experience} are more sensitive to the richness of information~\cite{oscar2017towards} in these interactions because they are typically more engaged in social communication.

% In addition, Table~\ref{tab:P_cor_eHMI-NV} also shows that Passengers' \textit{Agreeableness} scores had a significant negative partial correlation ($r=-0.251, p=0.014$), and  their \textit{Neuroticism} scores had a significant positive partial correlation ($r=0.214,p=0.037$) with their gaze duration on BD, respectively, when the APMV using the eHMI-AV. 

\subsection{Limitations}

The participants in this study are young Japanese, and the cultural and age factors may have influenced their specific gaze behavior.
With only 24 participants, the small sample size may limit the statistical power of the pairwise correlation results.  
The results may be limited by the specific eHMI designs used in this study.
The pedestrians in the experiment were played by experimenters.
Their behaviors and reactions in the interaction scenarios were not as varied as those of real pedestrians.
This limitation in diversity may potentially affect the results of this study.

%In this study, the reaction of the controlled according to our experiment design, i.e., a controlled laboratory study in the field. In the future study, more natural reactions of the pedestrian-APMV-passenger interaction could be included by video observation study using sensor, such as Gopro, lidar or depth camera to collect a richer and more natural dataset. 

\section{CONCLUSION}

This study investigated the differences in gazing behavior towards pedestrians' head by APMV passengers with different personalities when APMV communicates with pedestrians using three different types of eHMI
When using a visual-based eHMI, which caused passengers to struggle in perceiving the communication content, the results indicated that passengers with higher \textit{Neuroticism} scores, who were more sensitive to communication details, might seek cues from pedestrians' reactions.
In addition, eHMI-NV using neutral voice did not significantly affect the gaze behavior of passengers toward pedestrians, regardless of personality traits.
In contrast, eHMI-AV using affective voice encouraged passengers with high \textit{Openness to Experience} scores to focus on pedestrians' heads.

In summary, this study revealed how different eHMI designs influence passengers' gaze behavior and highlighted the effects of personality traits on their gaze patterns toward pedestrians, providing new insights for personalized eHMI designs.
In future research, we will further explore the essential causes of these effects.

\section*{ACKNOWLEDGMENTS}
This work was supported by JSPS KAKENHI Grant Number 22H00246, Japan.

\footnotesize
\bibliographystyle{IEEEtran}
\bibliography{sample.bib}

\end{document}